\documentclass[
 aps,
 pre,
 amsmath,amssymb,
 reprint,superscriptaddress,
showkeys
]{revtex4-2}

\usepackage{epigraph}
\usepackage[T1]{fontenc}
\usepackage[utf8]{inputenc}
\usepackage{amsmath}
\usepackage{amssymb}
\usepackage{graphicx}
\usepackage{refstyle}
\usepackage{skak}
\usepackage{multirow}
\usepackage{epsdice}
\usepackage{soul}
\usepackage[
  citecolor=blue,
  colorlinks,
  linkcolor=blue,
  urlcolor=blue,
]{hyperref}
\usepackage{epstopdf}
\usepackage{bm}
\usepackage[version=4]{mhchem}
\usepackage{tikz}
\usetikzlibrary{arrows,shapes,trees}
\usepackage{booktabs} 
\usepackage{lipsum} 

\begin{document}
	\title{Braess's paradox in tandem-running ants: When shortest path is not the quickest}
\author{Joy Das Bairagya}
\email{joydas@iitk.ac.in}
\affiliation{
 Department of Physics, Indian Institute of Technology Kanpur, Uttar Pradesh, PIN: 208016, India
}
\author{Udipta Chakraborti}
\email{udi1570@gmail.com}
\affiliation{Present address: Sorbonne Université, CNRS, Inserm, Neuro-SU, 75005, Paris, France
}
\affiliation{Sorbonne Université, CNRS, Inserm, Institut de Biologie Paris-Seine, IBPS, 75005, Paris, France
}
\affiliation{
Behaviour and Ecology Lab, Department of Biological Sciences, Indian Institute of Science Education and Research, Kolkata, Mohanpur 741246, India
}

\author{Sumana Annagiri}
\email{sumana@iiserkol.ac.in}
\affiliation{
Behaviour and Ecology Lab, Department of Biological Sciences, Indian Institute of Science Education and Research, Kolkata, Mohanpur 741246, India
}
\author{Sagar Chakraborty}
\email{sagarc@iitk.ac.in}
\affiliation{
 Department of Physics, Indian Institute of Technology Kanpur, Uttar Pradesh, PIN: 208016, India
}
\begin{abstract} 
Braess's paradox---where adding network capacity increases travel time---is typically attributed to selfish agents. Although eusocial colonies maximize collective fitness, we find experimentally that \emph{Diacamma indicum} ants exhibit this paradox: Leaders favour the shortest path even when it slows the colony. We present a quantitative model of the exploration-exploitation trade-off, demonstrating that evolutionary forces selecting for shortest-path identification can force suboptimal global states. This proves the paradox can emerge in highly cooperative systems without individual selfishness.
\end{abstract}
\maketitle
\section{Introduction}
Multiple paths often lead to a unique destination. For survival, if reaching the destination is critical, then finding the best path---minimising the travel time---becomes crucial. Intriguingly, even if selfish travellers possess all the information about the paths, they find it surprisingly hard to collectively agree on choosing the most beneficial path~\cite{Braess2005}. For social insects like ants, a crucial task for their survival is nest relocation: All colony members, including developing broods and stored resources, are relocated into a new habitable place. During transit multiple factors---such as predators~\cite{McGlynn2012}, weather conditions~\cite{McGlynn2012}, inhospitable environmental conditions~\cite{McGlynn2012}, and brood thefts~\cite{Paul2016}---directly impact the fitness consequences for all relocating colony members. Therefore, we would expect evolution to shape their ability to find the optimal path during nest relocation: The quicker the relocation, the lesser the amount of fitness sacrifices. Indeed, a recent study on the ant species, {\it Diacamma indicum}, has documented that most individuals use a path that takes the least amount of time to reach their new nest~\cite{Mukhopadhyay2019}.

Generally, the optimal quickest path is equivalent to the path which is shortest in length. However, the equivalence does not always hold when multiple individuals are trying to reach the same destination. A classic game-theoretic prototypical model of such a paradoxical situation is Braess' paradox~\cite{Braess2005,Banerjee2021,Sousa2013,Barbosa2014,Donovan2018,Coletta2016,Pala2012,Gounaris2024,Youn2008,lee2012exploring}, where the shorter the journey, the longer it takes when more than one individual tries to reach a destination simultaneously. Assuming that individuals are von-Neumann--Morgenstern (VNM) rational~\cite{VNMbook}, it can be shown that if everyone prefers the path with less travel time, then all select the path which is {shorter} in length. However, since everyone uses the same path leading to congestion, it is no longer the quickest path, but instead becomes a suboptimal one. The situation becomes a quagmire: When one individual unilaterally tries to choose an alternative path, she finds that all other paths take longer than the current situation; as VNM rationality demands individual selfishness, no one would deviate from their current situation unless they gain, and therefore, they remain stuck in the suboptimal situation.

It has been documented that ants perform better than humans in cooperative transport where multiple individuals are required to shift an object which is larger than their body size~\cite{Dreyer2024}. Surprisingly, it has also been documented that ants can successfully solve complex problems, such as the Towers of Hanoi~\cite{Reid2011}. Moreover, recent studies have shown that arboreal ants solve the problem of finding the shortest path in a very tangled network of tree branches to navigate between food and their nest~\cite{Garg2023}. Therefore, ants have been shown to solve complex optimization problems; and in nature, it is often observed as they navigate through intricate path networks like a network of tree branches, where Braess's paradox equivalent scenarios may naturally arise. 

Sometimes such a scenario may be a matter of life or death: A nest may need to be relocated by the ants to save the entire colony. Therefore, with respect to this task, we could expect them not to act as selfish, VNM rational agents. This cooperative nature of the ants during relocation raises an open and intriguing question: How do ants perform in a path network where VNM rational agents are known to fall prey to Braess's paradox? Given that rapid relocation is crucial for their fitness, it would be odd if evolution led the ants to behave selfishly. Instead, we expect them to choose a route that maximizes collective colony efficiency. Additionally, a recent experimental study has shown that the behaviour of {\it D. indicum} ant during relocation can be modelled using concepts of classical and behavioural game theories~\cite{DasBairagya2025}. Therefore, this study also helps to enhance the understanding of applicability of such game theoretic concepts in the context of ants.

Relocation of nest in {\it D. indicum} itself is a fascinating phenomenon, as they do not lay pheromone trail during relocation. Instead, they relocate their nest (all the nest mates, including all broods) from an uninhabitable place to a habitable one by tandem running. In a colony, a few  individual ants become leaders---the individuals who guide their nest-mates one by one to the new nest from the old nest. This method of transporting nest mates is called tandem running, where the leader is called the tandem leader (TL) and the nest mate who follows the leader is termed the follower. Together, they are labelled as a tandem pair. After reaching the new nest with a follower, the tandem leader again returns to the old nest to recruit a nest-mate; in this phase, the leader is termed the returning leader (RL). During this relocation, leaders make independent decisions to choose a path from the old nest to the new nest and vice versa. There is no colony-level communication present~\cite{Annagiri2023} to select a particular path; therefore, it is worth investigating whether their individual behaviour could emerge as an optimal behaviour for the whole colony if they face the situation of Braess's paradox, i.e., the shortest path no longer remains the quickest.

Technically, there is an additional catch: Traditionally, Braess's paradox is mostly presented in the context of unidirectional transport, whereas ant relocation is bidirectional---TLs move towards the new nest, and RLs move from the new to the old nest. Therefore, it is an interesting theoretical question whether Braess's paradox occurs in the case of bidirectional flow. To address this, let us first work with the theoretical version of bidirectional Braess's paradox; from intuition thus gained, we can build an experimental setup to examine the behavioural response of the tandem-running ants during relocation.
\begin{figure}
	\centering
	\includegraphics[width=0.9\linewidth]{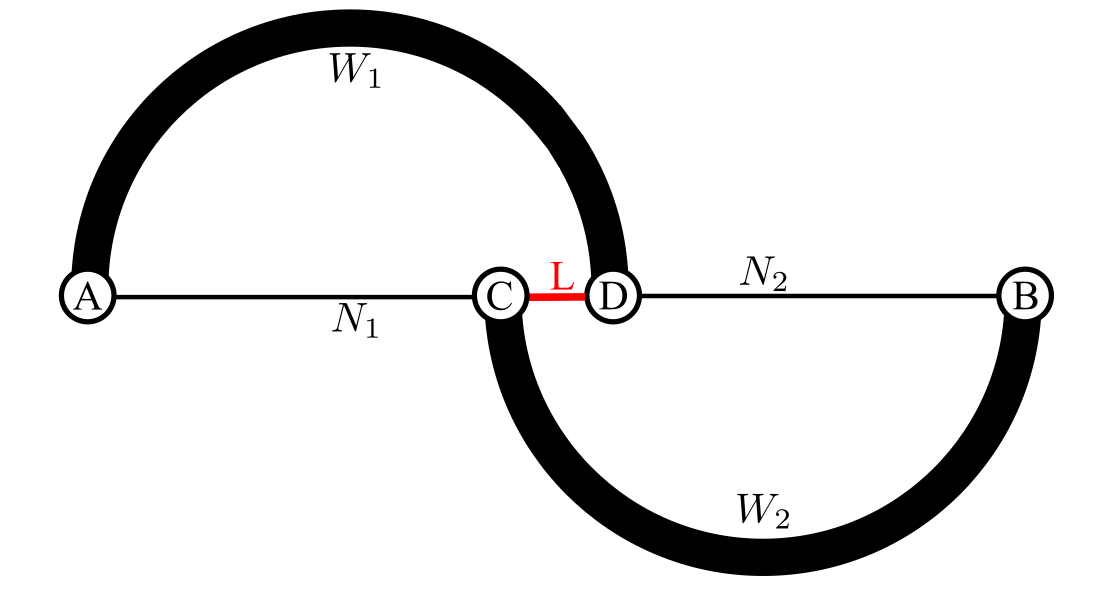}
	\caption{\emph{Schematic setup of Braess's paradox}: This figure shows a path network involving only two symmetric paths---$N_1W_2$ and $W_1N_2$---connecting origin, $A$, and destination, $B$, and  the link L, effecting almost zero travel time between $C$ and $D$.}
	\label{fig:fig1}
\end{figure}
\begin{figure*}
	\centering
	\includegraphics[width=1\linewidth]{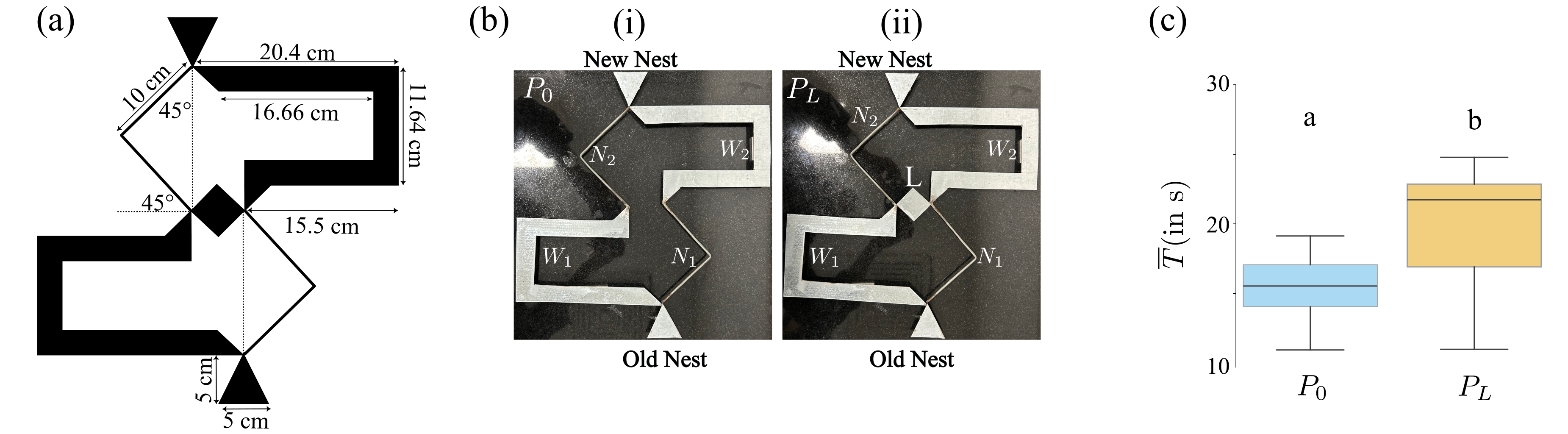}
	\caption{\emph{Braess's paradox in tandem-running ants}: Subfigure (a) is the experimental design of the path network used in the experiment, whereas subfigure (b) presents photos of the two path networks: $P_0$ (without the bridge)  and $P_L$ (with the bridge). Subfigure (c) shows that the relocation time per individual is statistically higher in $P_L$ than in $P_0$ (different Latin letters above the box-whisker plots indicate statistically significant differences). A Mann--Whitney $U$ test was performed for hypothesis testing with a significance level of $0.05$.}
	\label{fig:fig2}
\end{figure*}
\begin{figure}
\end{figure}

\section{Braess's paradox in bidirectional traffic}
The paradox can be understood using a minimal model with two path networks (see Fig.~\ref{fig:fig1}) and only two agents travelling in opposite directions. In the context of ant relocation, we can think of one agent as TL starting from old nest (say, A) and the other agent as the RL  starting from new nest (say, B); TL and RL want to reach the new nest and the old nest, respectively. Travel on identical long, wide path-segments $W_1$ and $W_2$ takes same time for two agents even if both of them are simultaneously on the same path. However, travel on identical short, narrow path-segments $N_1$ and $N_2$ takes more time when two agents are present simultaneously compared to when only one is on one path. This is simply because on narrow lanes the agents encounter each other and find hindrance in crossing each other. On the added path segment---linking bridge (L)---that facilitate travel in almost zero time for any agent on it: It essentially creates new complete paths, viz., $N_1N_2$ (combined path consisting of only narrow segments) and $W_1W_2$ (combined path consisting of only wide segments). Thus, the situation viewed as a game between two players (agents) such that each player have a set of strategies---$\mathcal{S}=\{N_1W_2, W_1N_2, N_1N_2, W_1W_2\}$---the paths to reach their destinations.

To determine the payoff structure of this game, we make the following assumption. The biological objective of tandem running is efficient nest relocation, and faster relocation enhances fitness. It is, therefore, reasonable to assume that the payoffs associated with each strategy are functions of the travel times of the players. Let $t^{ tl}_{\pi}$ and $t^{rl}_{\pi}$, respectively, denote the travel time of the TL and the RL to cover path segment $\pi \in \{N_1, W_1, N_2, W_2\}$. There are finite probabilities, say, $e_1$ and $e_2$, that the TL and  the RL encounter each other on $N_1$ and $N_2$, respectively. Due to such encounters, the TL and the RL, respectively, incur additional time costs $\delta^{tl}_{N_i}$ and $\delta^{rl}_{N_i}$ on a narrow path $N_i$ ($i\in\{1,2\}$). A payoff matrix of the game can now be easily constructed (see Appendix~\ref{uni_Braess's_paradox},~\ref{bi_Braess's_paradox}).

In the path network without the bridge, it is straightforward to verify that the game admits two pure NEs~\cite{Nash1950},
namely $(N_1W_2, W_1N_2)$ and $(W_1N_2, N_1W_2)$----strategy profiles in simultaneous play by the TL and the RL. In addition, the game also admits a mixed NE~\cite{Nash1950}, where each player randomizes over their two available strategies. When the bridge is put in, $(N_1N_2,N_1N_2)$ can turn out to be the NE if $1 < \frac{t^{k}_W - t^{k}_N}{e_N \delta^{k}_N} < \frac32$
for both $k=tl$ and $k=rl$ is satisfied. We assume, by construction, the times $t$ and $\delta$ are assumed to be independent of subscript $i$ associated with path segments and hence, the subscripts have been dropped. This NE is surprising as adopting it leads to the Braess's paradox: The players take more time to reach their destinations than when the bridge was absent.
 
We conclude that it is possible to build such networks where VNM rational~\cite{VNMbook} agents fall prey to Braess's paradox during their bidirectional movements. Thus, it would be very interesting to examine experimentally whether ants outperform the VNM rational agents in a situation like Braess's paradox.

However, modelling the behaviours using game theory during relocation of tandem-running \emph{D. indicum} colony raises a few issues which are very hard to resolve: First, in our toy model, we have used only two individuals whereas in the nest relocation, multiple ants are involved making the game difficult to theoretically analyze. Second, it is challenging to model their decision-making procedure mathematically, since ants should be realistically seen as procedural rational agents~\cite{Bairagya2025} than VNM rational. Third, it is even harder---probably near impossible---to write the game payoff structure because multiple individuals are involved and we can not infer {\it a priori} their preferences over multiple outcomes. Therefore, although game theoretic considerations showcase the paradox crystal clearly, we need to pragmatically shift from game theoretic analysis in order to theoretically comprehend the outcome observed in our experiment set up to study Braess's paradox. 

\section{Experimental Observations}
To bring the essence of Braess's paradox to the fore so that such a thing can be interpreted in a complex two-way traffic as in the case of relocation of ants, we first need to appreciate the crux of the paradox: The crucial factor is consideration of time, and the essence of the paradox is that the shortest path turns out not to be the quickest one. The reason of getting trapped into the paradox stems from the fact that decisions are being made at the individual level such that utilities (composed of benefit and cost) are also based on individual level success of completing that task of doing tandem runs to save the colony. While, of course, the task is undertaken because---evolutionarily speaking---ants are genetically hardwired to think about the collective benefit of the colony; the decision taken during a task at a time scale much smaller than evolutionary timescale is based on individual proximate reasons. If it was not so then ants would have collectively taken the quickest route and not the shortest one.

 We have carefully, after some educated trials and errors, constructed an experimental setup where the essence of the paradox can be exhibited even in the case of ants.  The diagram of experimental setup is presented in Fig.~\ref{fig:fig2}(a)---the rhomboidal bridge can be removed or added to get two path networks: $P_0$ and $P_L$, respectively. Further technical details about the setup are presented in Appendix~\ref{expt_details}. In the beginning, an entire ant colony is in the old nest and the ants move towards the new nest with some of them coming back to the old nest to lead other ants to the new nest, thereby creating a lot of bidirectional traffic encounters. In such a setup, the relevant physical parameter to qualitatively characterize Braess's paradox, would be $\overline{T}$---total transport time of the entire colony divided by the size of the colony---as it describes, on average, how much time a single ant takes to finally settle at new nest.

First, we took eight colonies of ants and disrupted their nest, thereby, forcing them to relocate to the new nest while presenting them with $P_0$ (Fig.~\ref{fig:fig2}b(i)) path network. Next, we took eight other colonies and initiated the similar relocations, but now with $P_L$ (Fig.~\ref{fig:fig2}b(ii)) as the available path network. What we observe (Fig.~\ref{fig:fig2}c) is that $\overline{T}$ is statistically larger---as per the Mann--Whitney $U$ test---in  $P_L$ compared to $P_0$; in other words, \emph{the ants do get trapped in Braess's paradox}!

\begin{figure}
	\centering
	\includegraphics[width=1\linewidth]{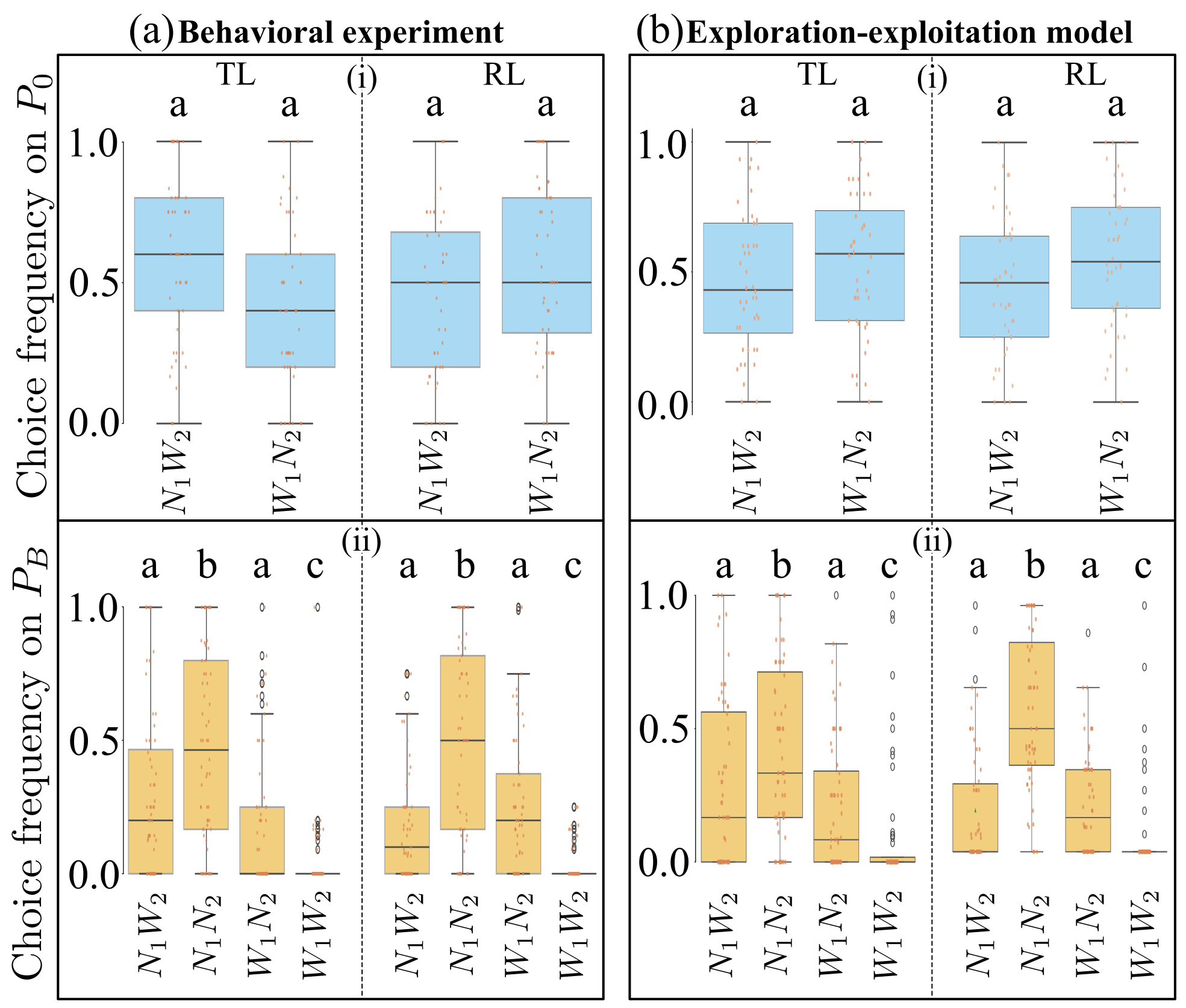}
	\caption{\emph{Exploration--exploitation paradigm explains Braess's paradox}: In subfigures (a) and (b), the path-choice-frequencies of leaders are shown using box-whisker plots, respectively, in experiments and in simulations. The subfigures tagged with (i) and (ii), respectively, denote that the corresponding plot correspond to path network $P_0$ and $P_L$. Different Latin letters above the box-whisker plots indicate statistically significant differences---same letters mean statistically similar. Here, we plot the choice-frequencies of 46 and 61 leaders---who individually performed at least four tandem runs and collectively accounted for more than 80\% of total tandem runs---across eight colonies for $P_0$ and $P_L$, respectively.  Statistical comparisons were performed using the  Wilcoxon test to compare choice frequencies across paths and the Mann--Whitney U test to compare experimental results with simulation outputs. In both cases, a significance level of $0.05$ was used.}
	\label{fig:fig3}
\end{figure}

This intriguing observation immediately raises a question: Which individual-level decisions during relocation lead ants to exhibit the paradox. To address this, we examine the route choices of all the leader ants (TLs and RLs)---who are responsible for transporting nest-mates from the old nest to the new nest. In the path network ($P_0$) without the bridge, we find using the Wilcoxon signed-rank test that frequencies of choice for paths $N_1W_2$ and $W_1N_2$ are statistically identical for both the TLs [Fig.~\ref{fig:fig3}a(i)] and for returning leaders [Fig.~\ref{fig:fig3}b(i)] Finding that the ants distribute themselves equally between the two available paths, might lead to conclusion that the ants select paths randomly, without any systematic bias. If this argument were to hold true in the path network ($P_L$) containing the bridge, one would expect identical decision frequencies across the four available paths: $N_1W_2$, $W_1N_2$, $N_1N_2$, and $W_1W_2$, analogous to the unbiased distribution observed in the network $P_0$. However, our observations reveal a markedly different pattern in the choice frequencies of both the TLs and the RLs: The leaders mostly prefer $N_1N_2$ path as depicted in Figs.~\ref{fig:fig3}b(ii). The per-individual transport time increases---leading to Braess's paradox---in the presence of the bridge: Rather than distributing themselves equally between the optimal paths $N_1W_2$ and $W_1N_2$, the ants preferentially choose the suboptimal path $N_1N_2$, which consists of narrow but shorter segments. This choice leads to increased congestion and, consequently, the ants require longer times to reach their respective destinations. 

\section{Theoretical Model}
In both $P_0$ and $P_L$, the observations curiously match closely with the predicted NEs in bidirectional Braess's paradox with only two agents; however, as argued earlier, it would be technically impossible to extend the game-theoretic arguments to the present real experiments. Nevertheless, as we emphasized earlier, we understand it is all about optimizing time spent to transport a colony to safety. We offer a simpler model to comprehend the experimental results: the paradigm of exploration-exploitation dilemma~\cite{Gueudr2014,Dichio2024,BlumMoyse2025}. The exploration-exploitation dilemma is a fundamental trade-off between searching for new possibilities (exploration) and leveraging known reliable options (exploitation) to maximize long-term gains. In the context of saving time, this balance is crucial because excessive exploration can waste time on fruitless searches, while over-exploitation can lock one into inefficient strategy. 

Effectively, we envisage a simplification where the multi-agent strategic interaction scenario may be seen as a kind of mean-field situation where each ant decides on a strategy irrespective what the other ants do---the presence of all other ants is effectively a background environment for the focal ant who mainly decides whether to exploit a known path or to explore new paths. We intuit that since during random exploration, the probability of discovery of new nest via the shortest path is maximum, the tendency to subsequently exploit would bias the path choice frequency towards the shortest path. 

To this end, we build a simplest yet non-trivial model of the process (technical details in Appendix~\ref{ex-ex}) wherein we assume that each leader initially explores all possible paths while performing a random walk. After discovering her destination (for the onward journey the destination is the new nest, while for the returning journey the destination is the old nest), she may exploit the discovered path rather than explore again, i.e., she chooses that path for the subsequent journey with probability $p$. 

We further assume that each leader in a colony performs the same number of
tandem runs and returning runs. Suppose that in the $j$-th colony there are $F_j$ followers---colony size minus number of leaders, $L_j$.
Therefore, each leader must accomplish $T_j = {\rm int}\left({F_j}/{L_j}\right)$ tandem runs and $T_j-1$ number of returning runs so that at the end of the process the leaders are finally in the new nest. Hence, each leader has $T_j$ opportunities to make a path choice either by exploration or exploitation. Let us assume that in her $k$-th opportunity, the leader chooses a path $s\in \{N_1W_2, W_1N_2, N_1N_2, W_1W_2\}$. The choice observation of the $i$-th leader in her $k$-th instance of choice as TL can therefore be represented by the four-dimensional vector  ${\bm \theta}_{ij}^{tl,k}=(\theta^{tl,k}_{N_1W_2}\,\,\, \theta^{tl,k}_{W_1N_2}\,\,\,\theta^{tl,k}_{N_1N_2}\,\,\,\theta^{tl,k}_{W_1W_2}),$ where $\theta^{tl,k}_{s}=1$ if the ant chooses the path $s\in\{N_1W_2, W_1N_2, N_1N_2, W_1W_2\}$ and $\theta^{tl,k}_{s}=0$ otherwise. Similarly, we introduce ${\bm \theta}_{ij}^{rl,k}$ corresponding to the RL. The choice frequency vectors of the $i$-th tandem and returning leader are then given by ${\bm x}^{tl}_{ij}=\frac{1}{T_j}\sum_{k=1}^{T_j}{\bm \theta}_{ij}^{tl,k}$ and ${\bm x}^{rl}_{ij}=\frac{1}{T_j-1}\sum_{k=1}^{T_j-1}{\bm \theta}_{ij}^{rl,k}$, respectively. 

Since ${{\bm x}}^{a}_{ij}$ ($a\in\{tl,rl\}$) is the average of multinomial observations, we approximate its distribution by a multivariate Gaussian, $f({{\bm x}}^{a}_{ij}|p)$, completely characterized by the mean vector of the choice frequency ${\bm \mu^a}$ and the covariance matrix ${\sf \Sigma}^a_j$ which depends on $p$. Assuming that each leader across all colonies acts as an independent decision maker, and that the onward and returning journeys are also independent, we maximize the log-likelihood of observing the choice frequency data [see Fig.~\ref{fig:fig3}a(i)-(ii)] under this probability distribution. Therefore, we maximize
	$
	\mathcal{L}(p)=
	\sum_{j=1}^{8}\sum_{i=1}^{L_j}
	\log\big[f({{\bm x}}^{tl}_{ij}|p)\big]+\log\big[f({{\bm x}}^{rl}_{ij}|p)\big],
	$
	and the maximum likelihood estimator of $p$ is given by
	$
	{p}_{\rm mle}=\arg\max_{p}\,\log \mathcal{L}(p).
	$
After obtaining the maximum likelihood estimate ${p}_{\rm mle}=0.690$ ($95\%$ confidence interval $[0.636, 0.747]$)---meaning, a leader ant is indeed biased towards exploitation, we performed an agent-based simulation (detailed in Appendix~\ref{ex-ex-ABM}) of the relocation process using eight colonies identical to those in the experiments (i.e., with the same colony sizes and number of leaders). We then verified that the distribution of path choice frequencies for both the onward and returning journeys is statistically indistinguishable with the experimental observations (see Fig.~\ref{fig:fig3}a(ii) and b(ii)), validating the effectiveness of the model.

\section{Conclusion}
Our work unambiguously establishes that Braess's paradox~\cite{Braess2005,Youn2008} can arise even in the absence of VNM rationality (individual selfishness)~\cite{VNMbook}. The individual decisions---made while being in Braess's paradox's trap---render the colony-level outcome worse and appears to contradict the cooperative nature of eusocial systems, thereby posing a theoretical puzzle. We show that within the context of tandem-running ants, the phenomenon can instead be explained through the exploration--exploitation paradigm~\cite{Chernoff1959,March1991,Biesmeijer2001}.  Evolutionary forces select a balance between exploration and exploitation such that tandem-running ants choose the path that is shortest in length, not the quickest one. The observation that they most often choose the path of shorter length aligns with earlier observations~\cite{Mukhopadhyay2019}. However, in the previous study, all the paths presented to ants during relocation were wide, allowing multiple ants to move together unhindered; whereas, in our study, the shortest path is congested for an ant to travel. Therefore, probably the observation of Braess's paradox in the ants may be viewed as an evolutionary spandrel~\cite{Gould1979, Rivoire2019}.

\section*{Data availability statement}
All data analyzed in this paper are available at Zenodo:~\href{https://doi.org/10.5281/zenodo.19044653}{https://doi.org/10.5281/zenodo.19044653}.\\
All codes used to generate the plot in this paper are available at Github:~\href{https://github.com/joydasbairagya/Braess-s-paradox-in-tandem-running-ants-When-shortest-path-is-not-the-quickest.git}{https://github.com/joydasbairagya/Braess-s-paradox-in-tandem-running-ants-When-shortest-path-is-not-the-quickest.git}. \\
\acknowledgements
J.D.B. gratefully acknowledges IIT Kanpur (India) for the financial support through Fellowship for Academic and Research Excellence (FARE).
\bibliography{Bairagya_ant}
\appendix

\section{Braess's paradox}
In this section we recall the classical Braess's paradox and then provide the mathematical details of the paradox possible in bidirectional traffic. These understandings were crucial in constructing the experimental setup as used in our research.
\subsection{Unidirectional traffic}
\label{uni_Braess's_paradox}
Let us begin by revisiting Braess’s paradox for rational agents in its simplest form. Suppose that $M = 40$ drivers wish to travel from point A to point B (see Fig.~\ref{fig:fig1}), and each driver behaves as a VNM-rational agent~\cite{VNMbook}. They may choose any of the available routes in the networks shown in Fig.~\ref{fig:fig1}. Assume that all drivers start simultaneously from A.
Furthermore, suppose that roads AC ($N_1$) and DB ($N_2$) are narrow in width and shorter in length relative to path AD ($W_1$) and CB ($W_2$). Naturally, travel time on the narrow roads ($N_1$ and $N_2$) increases as more individuals use them simultaneously.  For simplicity, we assume that the travel time on $N_1$ and $N_2$ is linearly proportional to the number of individuals, denoted by $m_{1}$ and $m_{2}$, respectively, taking those roads. We choose time units such that the travel time on $N_1$ is $t_{N_1}=m_{1}$ units, and on $N_2$ is $t_{N_2}=m_{2}$ units. In contrast, the wider roads $W_1$ and $W_2$ incur a fixed travel time, independent of traffic, which we set to $t_{W_1}=t_{W_2}=45$ units. Additionally, we assume that the linking bridge (L) connecting C and D [see Fig.~\ref{fig:fig1}] requires zero travel time $t_{L}=0$. Throughout, for an individual, we measure the utility of a road as the negative of its travel time; for example, the utility of road $N_1$ is $-t_{N_1}=-m_1$.

Now, let us determine the Nash equilibria (NE) in the  two path networks: without L and with L (see in Fig.~\ref{fig:fig1}). Since we assume that all drivers are VNM-rational, each driver is expected to select a Nash equilibrium route; thus, the aggregate distribution of drivers across the network must correspond to the NE. In the network without L, the NE is given by $m_{1} = m_{2} = \frac{M}{2} = 20$. No individual driver can improve her utility---i.e., reduce her travel time---{by unilaterally deviating from this configuration such that $m_{1} \neq m_{2}$. For instance, suppose a driver deviates from $N_1W_2$ to $W_1N_2$. Then the number of drivers on $W_1N_2$ becomes $21$, yielding a travel time of $21 + 45 = 66$ units, which is worse than remaining on $N_1W_2$, in which case the travel time would have been $20 + 45 = 65$ units.}

Similarly, the NE for the network with L occurs when all individuals take the path $N_1$, then the L, and then $N_2$ to travel from A to B; we call this path $N_1N_2$. However, this results in a travel time of $40 + 0 + 40 = 80$ units, which is strictly greater than the time (65) they would incur if they divided themselves equally between the two alternative routes---$N_1W_2$ and $W_1N_2$---without using the bridge. This demonstrates that rational individuals fail to achieve the collectively optimal outcome. This paradoxical situation, known as Braess’s paradox, is frequently observed in humans~\cite{Braess2005,Youn2008}, where adding an additional connection and thereby increasing the number of available options for $M$ independent VNM-rational agents ultimately worsens overall travel efficiency. 

\subsection{Bidirectional traffic}
\label{bi_Braess's_paradox}
Now we want to introduce the bidirectional Braess's paradox by extending the idea of unidirectional  paradox presented above in Appendix.~\ref{uni_Braess's_paradox}.  However, for simplicity and analytical tractability, here we consider the system to have only two strategically interacting individuals. Let us assume, for concreteness, that the two players of the game are the tandem leader (TL) and the returning leader (RL). The strategy set of a player, when linking bridge L is not present, is denoted by $\mathcal{S}_i$ where
\[
\mathcal{S}_i=\{N_1W_2,\; W_1N_2\};
\]
$i$ is put as either $tl$ or $rl$ depending on whether the player is TL or RL, respectively. Each strategy corresponds to a choice of path used to reach the respective destination. For instance, the strategy $N_1W_2$ indicates that an individual traverses both path segments $N_1$ and $W_2$. In particular, if the RL adopts the strategy $N_1W_2$, this means that she uses the segment $W_2$ followed by $N_1$ to reach the old nest from the new nest. Also, recalling the notation $\pi \in {\Pi}\equiv\{N_1, W_1, N_2, W_2\}$, we note that a path from A to B can be denoted as $\pi_j\pi_k$ ($j,k\in\{1,2\}$; $j\ne k$); in what follows, an element of $\mathcal{S}_i$ can be equivalently seen as a two-element subset of $\Pi$.

If the TL and the RL adopt strategies $s_{tl} \in \mathcal{S}_{{tl}}$ and $s_{rl} \in \mathcal{S}_{{rl}}$, respectively, 
their respective payoffs are given by
\begin{align}
	{\sf U}_{tl}(s_{tl},s_{rl})
	&= - \sum_{\pi \in s_{tl}} t^{tl}_\pi
	- \sum_{\pi' \in s_{tl} \cap s_{rl}} e_{\pi'} \, \delta^{tl}_{\pi'}, \\
	{\sf U}_{rl}(s_{tl},s_{rl})
	&= - \sum_{\pi \in s_{rl}} t^{rl}_\pi
	- \sum_{\pi' \in s_{tl} \cap s_{rl}} e_{\pi'} \, \delta^{rl}_{\pi'} .
\end{align}
Here, the intersection $s_{tl} \cap s_{rl}$ denotes the set of path segments that are common to both the TL and the RL. For example, if both players choose the strategy $N_1W_2$, then the common path segments are $N_1$ and $W_2$. Additionally, for the wide paths $W_1$ and $W_2$, we assume that no additional cost is involved if they encounter in a wide path, i.e., $
\delta^{tl}_{W_1}=\delta^{tl}_{W_2}=\delta^{rl}_{W_1}=\delta^{rl}_{W_2}=0.$

Since the path segments $N_1$ and $N_2$ are identical in both length and width, we assume that the travel times on these paths are the same for tandem pairs and returning leaders, i.e., $t^{tl}_{N_1} = t^{tl}_{N_2} = t^{tl}_{N}$ and $t^{rl}_{N_1} = t^{rl}_{N_2}=t^{rl}_{N}$. By the same reasoning, for the wide paths $W_1$ and $W_2$, we assume $t^{tl}_{W_1} = t^{tl}_{W_2}=t^{tl}_{W}$ and $t^{rl}_{W_1} = t^{rl}_{W_2}=t^{rl}_{W}$. Moreover, we assume that the probabilities of encounter on the narrow paths are identical, i.e., $e_{N_1} = e_{N_2}=e_{N}$. Finally, we assume that the additional time costs due to an encounter on the narrow paths are the same for both players:
$\delta^{tl}_{N_1} = \delta^{tl}_{N_2}=\delta^{tl}_{N}$ and 
$\delta^{rl}_{N_1} = \delta^{rl}_{N_2} = \delta^{rl}_{N}$.

Under these assumptions, we determine the Nash equilibrium (NE) strategies $(\hat{s}_{tl}, \hat{s}_{rl})$, defined as a pair of strategies for the tandem pair and the returning leader that satisfy
\begin{align}
	{\sf U}_{tl}(\hat{s}_{tl},s_{rl}) 
	&\geq {\sf U}_{tl}(s_{tl},s_{rl}), \\
	{\sf U}_{rl}(s_{tl},\hat{s}_{rl}) 
	&\geq {\sf U}_{rl}(s_{tl},s_{rl}),
\end{align}
$\forall$~$s_{tl} \in \mathcal{S}_{tl}$ and $s_{rl} \in \mathcal{S}_{rl}$. It is straightforward to verify that the game admits two pure strict NEs,
namely $(N_1W_2, W_1N_2)$ and $(W_1N_2, N_1W_2)$. In both of these NEs, the total travel time for $tl$ and $rl$ is identical:
$t^{tl}_{\rm NE} = t^{tl}_N + t^{tl}_W$ and $t^{rl}_{\rm NE} = t^{rl}_N + t^{rl}_W$. In addition, the game also admits a mixed NE, where each player randomizes over their available strategies. 

To determine the mixed NE, let the probability distribution over the two available strategies of a tandem leader be
\[
{\bf X}_{tl} =
\begin{pmatrix}
	x^{{tl}}_{N_1W_2} \\
	x^{{tl}}_{W_1N_2}
\end{pmatrix},
\]
and similarly, for the returning leader,
\[
{\bf X}_{rl} =
\begin{pmatrix}
	x^{{rl}}_{N_1W_2} \\
	x^{{rl}}_{W_1N_2}
\end{pmatrix}.
\]
The mixed NE $({\bf \hat{X}}_{tl}, {\bf \hat{X}}_{rl})$ is then defined by
\begin{align}
	{\sf U}_{tl}({\bf \hat{X}}_{tl}, {\bf X}_{rl}) 
	&\geq {\sf U}_{tl}({\bf X}_{tl}, {\bf X}_{rl}), \\
	{\sf U}_{rl}({\bf X}_{tl}, {\bf \hat{X}}_{rl}) 
	&\geq {\sf U}_{rl}({\bf X}_{tl}, {\bf X}_{rl}),
\end{align}
for all ${\bf X}_k$ satisfying $0 \leq x^k_{N_1W_2} \leq 1$, 
$0 \leq x^k_{W_1N_2} \leq 1$, and 
$x^k_{N_1W_2} + x^k_{W_1N_2} = 1$ where $k \in \{{tl}, {rl}\}$.

For the present case, the mixed NE is given by
\[
{\bf \hat{X}}_{tl}=
\begin{pmatrix}
	\frac{1}{2} \\
	\\
	\frac{1}{2}
\end{pmatrix},
\qquad
{\bf \hat{X}}_{rl} =
\begin{pmatrix}
	\frac{1}{2} \\
	\\
	\frac{1}{2}
\end{pmatrix}.
\]
Physically, this corresponds to a situation in which both the tandem pair and the returning leader choose each of their available paths with equal probability $\frac{1}{2}$. Unlike the two pure Nash equilibria, in the mixed Nash equilibrium the probability that TL and RL  {both sharing the same narrow path segment} is non-zero; in fact, it is $\frac{1}{2}$. Therefore, the mixed equilibrium strategy is more time-costly than the pure equilibria. The corresponding time costs for the TL and the RL are $t^{tl}_{\rm MNE} = t^{tl}_N + t^{tl}_W + \frac{1}{2}e_N\delta^{tl}_N$ and $t^{rl}_{\rm MNE} = t^{rl}_N + t^{rl}_W + \frac{1}{2}e_N\delta^{rl}_N$, respectively.

Now let us analyse how the above-described game changes when a linking bridge (L) is introduced  (see Fig.~\ref{fig:fig1}). By design, L creates two additional paths that allow each player to reach their destination. The two new paths are $W_1W_2$ and $N_1N_2$.

Consequently, the strategy sets of the TL and RL expand to four possible paths,
\[
\mathcal{S}'_{i} = \{N_1W_2,\, N_1N_2,\, W_1N_2,\, W_1W_2\};
\]
among these, a subset consisting of two newly available paths requires the use of bridge:
\[
\mathcal{S}^L_{i} = \{N_1N_2,\, W_1W_2\}.
\]

We now write the modified payoff functions for this new path network. When the TL and the RL adopt strategies $s'_{tl} \in \mathcal{S}'_{tl}$ and $s'_{rl} \in \mathcal{S}'_{rl}$, respectively, their payoffs are given by
\begin{align}
	{\sf U}_{tl}(s'_{tl},s'_{rl})
	&= - \sum_{\pi \in s'_{tl}} t^{tl}_\pi
	- \sum_{\pi' \in s'_{tl} \cap s'_{rl}} e_{\pi'} \, \delta^{tl}_{\pi'}
	- \chi^{{tl}}_L \, t^{{tl}}_L, \\
	{\sf U}_{rl}(s'_{tl},s'_{rl})
	&= - \sum_{\pi \in s'_{rl}} t^{rl}_\pi
	- \sum_{\pi' \in s'_{tl} \cap s'_{rl}} e_{\pi'} \, \delta^{rl}_{\pi'}
	- \chi^{{rl}}_L \, t^{{rl}}_L.
\end{align}
Here, $t^{{tl}}_L$ and $t^{{rl}}_L$ denote the time required to cross the bridge by the TL and the RL, respectively. The indicator functions $\chi^{{tl}}_L$ and $\chi^{{rl}}_L$ take non-zero value (specifically, $1$) if $s'_{tl} \in \mathcal{S}^L_{tl}$ and $s'_{rl} \in \mathcal{S}^L_{rl}$, respectively.

Analogous to the analysis done in the case of unidirectional traffic, in the presence of bidirectional traffic, Braess's paradox would be realized if $(N_1N_2,\, N_1N_2)$ happens to be the NE although the travel time experienced at this equilibrium exceeds that of the socially optimal configuration, viz., the NEs of the path network without the bridge L ($P_0$). Mathematically, we find that Braess's paradox is realized if
\begin{eqnarray}
	&0 < t^{{i}}_W - t^{{i}}_N - t^{{i}}_L - e_N \delta^{{i}}_N < \frac12 e_N \delta^{{i}}_N,\,\forall i\in\{tl,rl\}.\qquad
\end{eqnarray}
The first inequality results from application of definition of strict NE, whereas the second inequality results from the condition that time taken ($2t^i_N+t^i_L+2e_N\delta^i_N$) on $N_1N_2$ is more than time taken ($t^i_{\rm MNE}$) in mixed NE on $P_0$. {In the main text, we have used $t^i_L=0$. }

\section{Experimental Protocols}
\label{expt_details}
\begin{figure}
	\centering
	\includegraphics[width=1\linewidth]{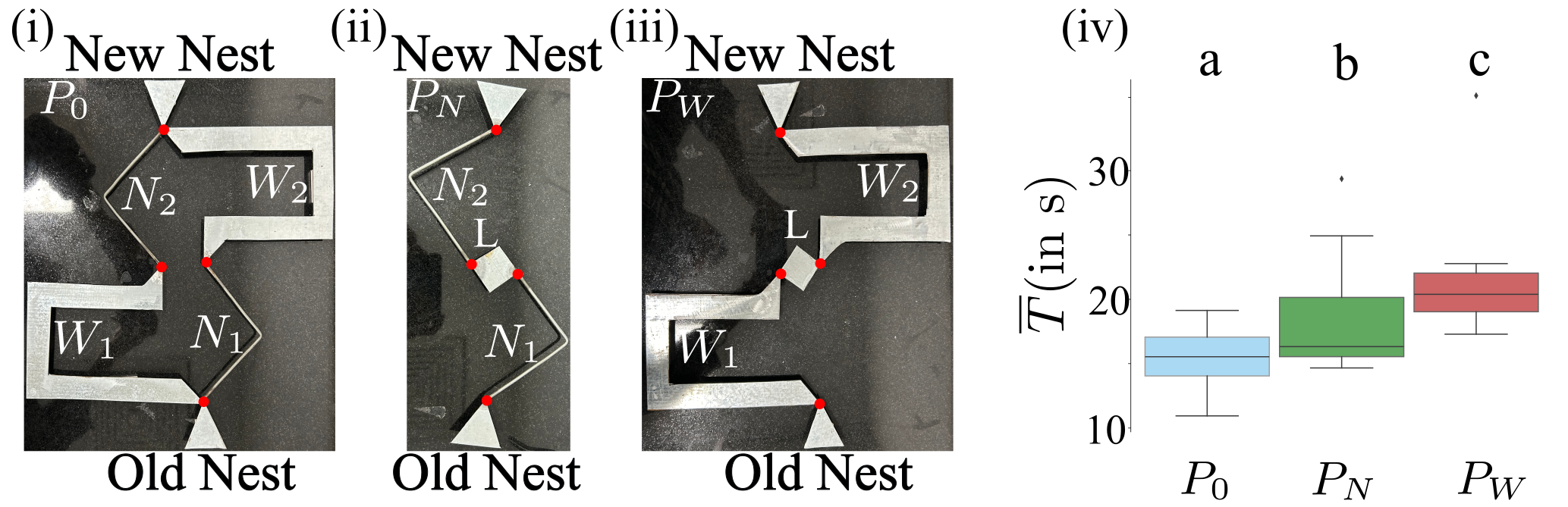}
	\caption{\textbf{Experimental setup for Braess's paradox grounded on the unidirectional Braess's paradox.}
		(i) Path network $P_0$ used as the control setup without a bridging link for the relocation experiments. 
		(ii) Path network $P_N$, where two narrow segments are connected by the linking bridge $L$. 
		(iii) Path network $P_W$, where two wide segments are connected by the bridge $L$. 
		(iv) Distribution of $\overline{T}$ (total relocation time per ant)  for colonies relocating through the three path networks. 
		The mean relocation time on path $P_0$ is significantly lower than that on $P_N$ and $P_W$, while the relocation time on $P_N$ is lower than that on $P_W$. 
		Thus, the ordering of relocation times satisfies the requirement for observing the unidirectional Braess's paradox, motivating the use of this setup in our experiments. 
		Statistical comparisons were performed using the Wilcoxon signed-rank test (significance level $0.05$), as repeated relocations were conducted using the same colony.}
	\label{fig:app4}
\end{figure}
\label{app:experiments}
We collected a total of sixteen colonies of \textit{Diacamma indicum} (Santschi, 1920~\cite{santschi1920cinq}) from the IISER Kolkata campus at Mohanpur, Nadia, West Bengal, India (22°56'N, 88°31'E) between November 2024 and February 2025. Each colony was collected from the field using the nest-flooding method~\cite{Kaur2012}. Following collection, all colonies were transported to the laboratory and maintained according to standard protocols~\cite{Mukhopadhyay2019}. Colonies were provided \textit{ad libitum} food (ant cake) and water on a regular basis~\cite{Bhatkar1970}. Only colonies containing a gamergate (the sole reproductive female) were used in the experiments. For each colony, we recorded the total number of adult females, adult males, and brood items (eggs, larvae, and pupae). To uniquely identify individuals within a colony, all ants were marked using non-toxic enamel paint (Testors, Rockford, IL, USA).

All relocation paths were made of stainless steel to eliminate any residual odour that could influence the natural decision-making of the ants. Furthermore, at each decision point, the available options (i.e., paths) were arranged symmetrically to prevent any directional bias arising from their spatial positions. Additionally, the relocation paths were placed inside a glass arena filled with water and kept in a closed room. The paths were positioned such that they remained dry, while the surrounding water prevented ants from leaving the designated paths. To avoid any global directional cues, the entire relocation setup was rotated to different orientations inside the water-filled arena for every relocation experiments.

Constructing the experimental set-up was challenging because, \textit{a priori}, it was not clear what type of path network and what distances of the narrow and wide path segments would be appropriate to test Braess's paradox. To guide the design, we drew inspiration from the classical one-directional Braess's paradox. However, in that formulation all individuals are assumed to be identical and traverse paths in equal time. Therefore, our goal was to construct a path network in which the core element of the paradox remains valid when measured using the average relocation time ($\overline{T}$). A crucial requirement of such a structure is the presence of two types of path elements: a narrow but short path ($N_1$ and $N_2$) [see Fig.~\ref{fig:app4} (ii)]and a wide but long path ($W_1$ and $W_2$) [see Fig.~\ref{fig:app4} (iii)]. Only narrow path network ($P_N$) should allow a single individual to traverse it faster than the wide path network ($P_W$), but if multiple individuals attempt to use it simultaneously, congestion should increase the travel time (see Fig.~\ref{fig:app_1}). At the same time, the segment lengths must be chosen such that if all individuals use the narrow paths, the per-individual travel time remains lower than if they all use the wide but longer paths~Fig.~\ref{fig:app4}~(iv). Furthermore, when a symmetric combination of narrow and wide paths ($P_0)$) is introduced~Fig.~\ref{fig:app4}(i), the resulting travel time should be lower than that obtained when individuals exclusively use the narrow paths~Fig.~\ref{fig:app4} (iv). To satisfy these requirements, we constructed three path configurations: one consisting only of narrow paths, one consisting only of wide paths, and a third consisting of symmetrically arranged narrow and wide paths. Through trial and error we determined the lengths of the narrow and wide path segments such that the above conditions were statistically satisfied. The final geometry of the path network that we use in our experiments is shown in Fig.~\ref{fig:fig2} (a), and the statistical verification of the aforementioned conditions is presented in Fig.~\ref{fig:app4}(iv).

For experiments conducted on the path network without the linking bridge, $L$ [Fig.~\ref{fig:app4}(i)], we used eight colonies with an average of $93.9 \pm 9.4$ (mean $\pm$ SD) adult female ants. The same colonies were used to conduct relocation experiments on the {$N_1N_2$ and $W_1W_2$} paths [Fig.~\ref{fig:app4}(ii)-(iii)]. After each relocation event, colonies were allowed to rest for at least 24 hours before being subjected to another relocation on a different path. In the path network containing the bridge $L$ [Fig.~\ref{fig:fig2} b (ii)], we used a separate set of eight colonies, distinct from those used in the no-bridge experiments. These colonies had an average of $98.1 \pm 11.6$ (mean $\pm$ SD) adult female members. Colony sizes used in the two experimental conditions were statistically comparable (Mann--Whitney $U = 24$, $p$-value = 0.56). 

\begin{figure}
	\centering
	\includegraphics[width=1.0\linewidth]{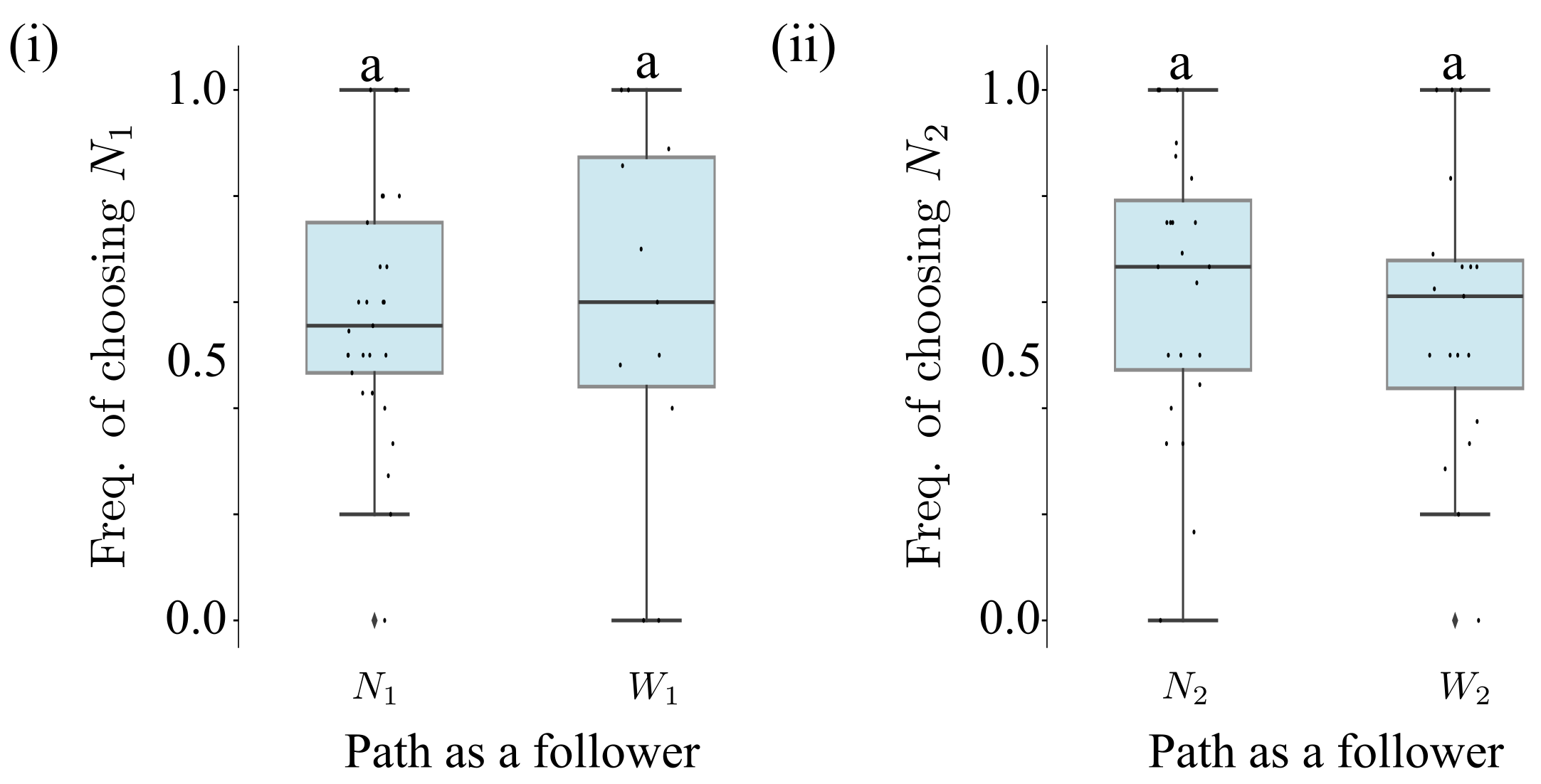}
	\caption{Unbiased exploration by secondary leaders on $P_0$ path network: 
		Subfigure (i) compares (using Wilcoxon signed-rank test, with significance level $0.05$) the following two frequencies of choosing $N_1$ path segment by the secondary leaders during the entire transport phase: frequency corresponding to the secondary leaders who, as followers, were relocated to new nest via $N_1$ and frequency corresponding to the secondary leaders who, as followers, were relocated to new nest via $W_1$. The similar subfigure (ii) corresponds to secondary leaders taking $N_2$ and $W_2$ path segments as followers. Subfigures unequivocally indicate that the path chosen by secondary leaders is independent of the path through which they were previously guided as followers by TLs. The box-whisker plots summarize the distribution of choice frequencies of 45 secondary leaders.}
	\label{fig:fig7}
\end{figure}

\begin{figure*}
	\centering
	\includegraphics[width=1.0\linewidth]{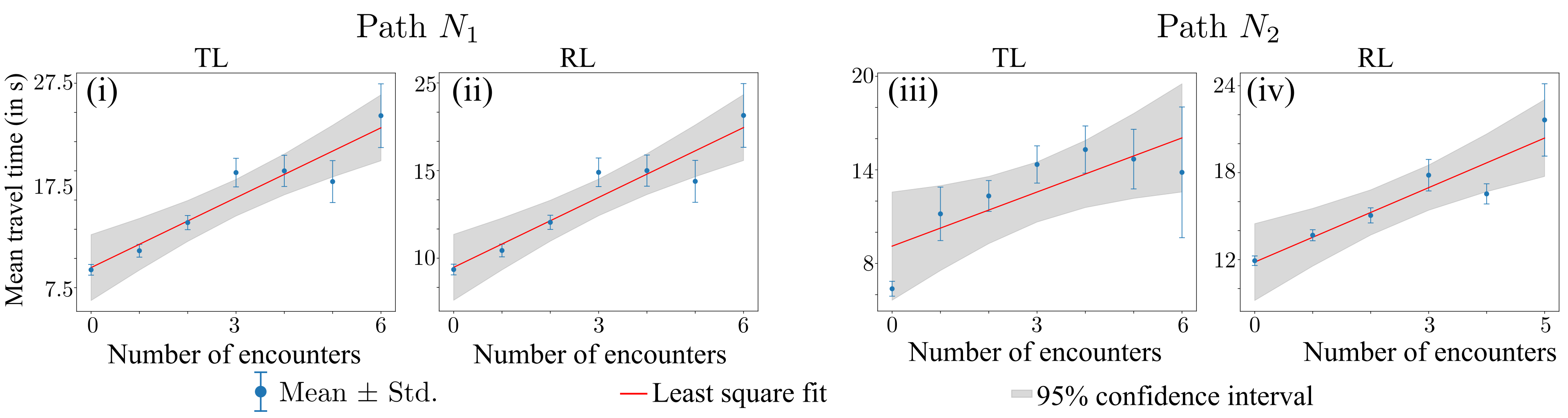}
	\caption{Encounters on the narrow paths delay a leader's journey: The four subfigures depict how mean travel time ($y$-axis) of a leader---TL (subfigures i and iii) or RL (subfigures ii and iv)---positively correlates with number of encounters ($x$-axis) she faced on paths $N_1$ (subfigures i and ii) and $N_2$ (subfigures iii and iv). A mean travel time at every fixed number of encounter has been calculated using all the journeys where corresponding leaders faced that fixed number of encounters. The red line represents the least squares fit, whereas the grey region is the $95\%$ confidence interval. Additionally, for each point corresponding to the mean, the standard deviation is shown using a vertical line. }
	\label{fig:app_1}
\end{figure*}

In all colonies, the number of pupae was very small relative to the colony size (never exceeding three). Therefore, for all calculations---both in experiments and in theoretical model, we neglect the pupae count and consider only total number of ants as the colony size. Another aspect of these ants is worth highlighting: Technically, there are two types of leaders---\textit{primary leaders} who never become followers before initiating a tandem run, and \textit{secondary leaders}~\cite{Franks2006,Mglich1974} who follow primary leaders before initiating their own tandem run. In the present work, however, we need not bother about this distinction because our experiments indicate that there is no effect of the path a secondary leader had previously taken as a follower on her subsequent path-choice-frequency during her tandem-run as a leader (Fig.~\ref{fig:fig7}). In other words, both kinds of leader are equivalent in our present study.

Prior to each experiment, all paths were cleaned with ethanol to eliminate any residual chemical cues from previous relocations. Subsequently, a colony along with its old nest (a Petri dish covered with red cellophane) was gently placed at one end of the path, while the new nest (also a Petri dish with a red cellophane cover) was positioned at the opposite end. Colonies were left undisturbed for 15 minutes to acclimate to the experimental environment. To initiate relocation, the roof of the old nest was carefully removed, and a white light source was positioned 15 cm above the nest to induce nest abandonment.

Ants were then allowed to relocate naturally from the old nest to the new nest. We ensure that every member of the colony visit the new nest at least one time. We define the \textit{transport phase} as the period from the first tandem run until the completion of transport, by which time every colony member has visited the new nest at least once. The total duration of this phase is referred to as the \textit{transport time}. Experiments on the path network without the link were recorded using two video cameras (Panasonic HC-V270), each focused on one nest entrance, at a frame rate of 25 frames per second (fps). In contrast, experiments on the path network with the link were recorded using a single video camera (Panasonic HC-V270) covering only the linking bridge (L) as the decision of the path choices can easily visible from the link L.


\section{Exploration-exploitation paradigm}
\label{ex-ex}
We aim to apply the exploration-exploitation paradigm to understand the individual behaviours observed in the experiment, as shown in Fig.~\ref{fig:fig3}. First, we present the numerical simulations of an agent-based model (ABM)---which will also be investigated analytically in the subsequent subsection---to examine the choice frequencies of individual leaders. 

\subsection{Agent-based model}
\label{ex-ex-ABM}
An agent has two properties: it may explore or exploit a path. At the start of the ABM, all agents begin to explore with probability 1; no one starts to exploit initially. After successfully reaching the destination once, it can begin to exploit the same path through which it has discovered the destination. Otherwise, it may restart exploring all the routes. For the TL, the destination is the new nest, and for the RL, the destination is the old nest.

First, let's explain the process of exploration. Primary TLs are known to discover the new nest from the old nest by exploring the available paths~\cite{Mukhopadhyay2019}. Additionally, we have assumed that the path used during the return journey is independently discovered by the leader during exploration from the new to the old nest. As a result, the paths used during the onward and the return journeys are allowed to differ, which is consistent with repeated experimental observations (see Fig.~\ref{fig:fig3}). Moreover, we have assumed the exploration of an agent does not depend on the exploration or exploitation of the other agents, i.e., no strategic interactions among agents have been considered. Moreover, experimental evidence shows that there is no statistically significant correlation between the path a secondary leader uses in her first tandem run and the path through which she is led as a follower to the new nest [see Fig.~\ref{fig:fig7}]. Accordingly, in our ABM, we assume that all leaders, both primary and secondary, independently explore the path network and select a path based solely on their own exploration experience.

\subsubsection{Exploration in $P_0$}
\label{withoutL}
\begin{figure*}
	\centering
	\includegraphics[width=1.0\linewidth]{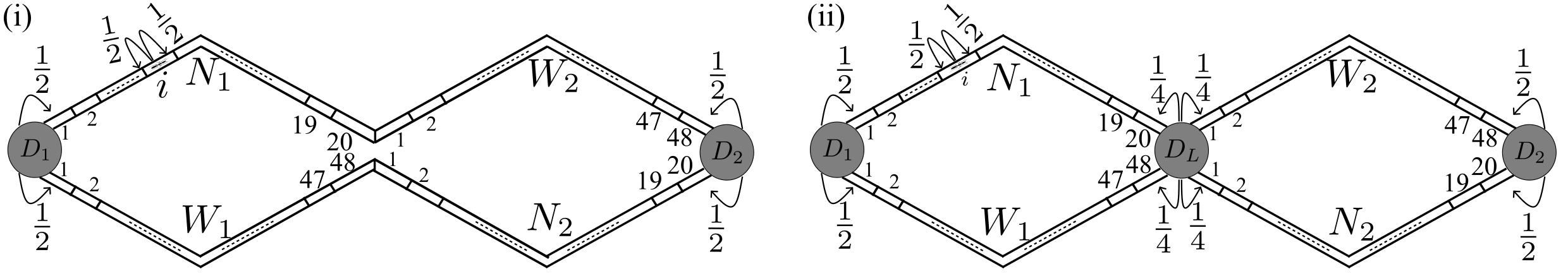}
	\caption{
		Schematic of the agent-based model (ABM). 
		(i) Structure of the ABM used to simulate exploration-exploitation dynamics on path network, $P_0$. The network consists of four path segments, $N_1$, $N_2$, $W_1$, and $W_2$. The segments $N_1$ and $N_2$ each contain 20 steps, while $W_1$ and $W_2$ each contain 48 steps, consistent with the relocation experiment. Two decision points, $D_1$ and $D_2$, represent the junctions near the old nest and the new nest (triangular regions in Fig.~\ref{fig:fig2}b). 
		(ii) Structure of the ABM for path network $P_L$, incorporating exploration-exploitation dynamics with an additional decision point $D_L$, which represents the linking bridge (see Fig.~\ref{fig:fig2}b(ii)). The arrows, with $\frac12$ and $\frac14$ written overhead, represent the probability of jump to the nearby cells.
	}
	\label{fig:abm}
\end{figure*}

The exploration phase in the ABM is modelled as an unbiased random walk [see Fig.~\ref{fig:abm} (i)]. We assume that interactions among agents do not affect the exploration phase. The lengths of the paths are chosen to preserve the same ratios as those used in the experiments. In the ABM, the lengths of paths $N_1$ and $N_2$ are set equal to $l_N=20$ units, while those of $W_1$ and $W_2$ are set equal to $l_W=48$ units. {Using an agent-based model (ABM), we aim to calculate the probability of selecting a particular path from an origin to a destination. In the real experiment with $P_0$ path, there are two decision points where multiple paths meet, and a leader must choose a path. These are: decision point $D_1$, where paths $N_1$ and $W_1$ connect and $D_2$, where $N_2$ and $W_2$ connect. In the ABM, we assume that $L$ is itself a decision point that connects all four paths $N_1$, $W_1$, $N_2$, and $W_2$ [see Fig.~\ref{fig:abm} (i)]. Furthermore, the ABM considers only the lengths of the paths between decision points, and ignores the distances from the old nest to $D_1$ and from $D_2$ to the new nest. Since the probability of reaching $D_1$ from the old nest and reaching the new nest from $D_2$ (and vice versa) is unity, these additional distances do not affect the probability of path selection. Therefore, we consider only the path lengths between the decision points $D_1$ and $D_2$.}

During exploration, a forward step corresponds to movement of one unit length toward the destination nest, whereas a backward step corresponds to movement of one unit length toward the origin nest. For an agent, the probability of taking a forward step and a backward step is $\frac{1}{2}$ in one unit of time~[see Fig.~\ref{fig:abm} (i) and (ii)]. Moreover, we assume that all available paths at a given decision point are chosen with equal probability [see Fig.~\ref{fig:abm} (i)]. This ABM framework, therefore, yields intrinsic path preferences that arise from an unbiased random exploratory dynamics. 

The dynamics for an individual agent during the onward journey are implemented as follows:
\begin{enumerate}
	\item The agent starts from decision point $D_1$ [see Fig.~\ref{fig:abm} (i)].
	\item From point $D_1$, the agent randomly selects one of the two outgoing path segments, $N_1$ or $W_1$, with equal probability, as observed in behavioural experiments~[see Fig.~\ref{fig:abm} (i)].
	\item After selecting a path, the agent moves one step forward or backward along that path with equal probability, corresponding to an unbiased random walk.
	\item If the agent reaches the point $D_2$, we record the path ultimately used to travel from point $D_1$ to point $D_2$.
\end{enumerate}
The same protocol is used to model exploration during the return journey, except that the agent starts from the point $D_2$ and terminates upon reaching the point $D_1$.

Now, let us discuss the exploitation that occurs during the tandem run and the returning journey, where a leader uses a path that she has discovered. Suppose in a colony $F$ number of followers are present and there are $L$ leaders (no secondary leaders). If we assume that every individual does the same number of tandem runs and returning journeys, then one leader has to do $T=\mathrm{int}\left(\frac{F}{L}\right)$ number of onward and $T-1$ returning journeys. Therefore, there are a total of $T$ and $T-1$ choices for an agent during forward journey and returning journey, respectively. In each of these choices, a leader may choose a path that she has discovered during exploration with probability $p$; otherwise, she starts exploring again. 

We have calculated the frequencies of the agents with which it chooses a path out of its $T$ and $T-1$ choices during tandem run and returning journey. In the experiment, we used eight colonies of different sizes. Therefore, in the ABM, we have also considered eight colonies of the same size as those used in the experiment. Based on experimental observations, we have determined the number of leaders who have completed at least four tandem runs for each colony; the same number of leaders is used in the ABM. Therefore, we have used eight different $F$ and $L$ that we determined from the eight ant colonies used in the experiment. Using this model, we find that the preferences for the paths $N_1W_2$ and $W_1N_2$ during tandem running are statistically indistinguishable [Fig.~\ref{fig:fig3}b (i)]. A similar symmetry is observed during the return journey, where the paths $W_2N_1$ and $N_2W_1$ are chosen with equal probability [Fig.~\ref{fig:fig3}b(i)].
\subsubsection{Exploration in $P_L$}
Next, we examine exploration-driven preferences under the same assumptions when a bridge L, i.e., a decision point $D_L$ is introduced [see Fig.~\ref{fig:abm}(ii)]. {In the  relocation experiment on path $P_L$, there are four decision points. These are: decision point $D_1$, where paths $N_1$ and $W_1$ connect; $D'_L$, where $N_1$, $W_2$, and $L$ connect; $D''_L$, where $W_1$, $N_2$, and $L$ connect; and $D_2$, where $N_2$ and $W_2$ connect. In the ABM, we assume that $L$ is itself a decision point $D_L$ that connects all four paths $N_1$, $W_1$, $N_2$, and $W_2$ [see Fig.~\ref{fig:abm}(ii)].}  In this case, the ABM protocol for the onward journey is modified as follows:
\begin{enumerate}
	\item The agent starts from the point $D_1$.
	\item From point $D_1$, the agent randomly selects one of the two outgoing paths, $N_1$ or $W_1$, with equal probability.
	\item The agent performs an unbiased random walk along the chosen path.
	\item Upon reaching the point $D_L$,  she selects one of the four paths $N_1$, $N_2$, $W_1$, or $W_2$ with equal probability.
	\item After selecting a path, the agent again performs an unbiased random walk.
	\item If the agent reaches the point $D_2$, we record the path used to reach point $D_2$ from $D_1$.
\end{enumerate}
Analogous procedure is followed for the return journey, with the agent starting from the point $D_2$. Moreover, the exploitation phase is same as that we have used for the case without the bridge (see in the Sec.~\ref{withoutL}).

Using this ABM, we determine how the introduction of a bridge $D_L$ alters path preferences. During the onward journey from the old nest to the new nest, leaders most frequently choose the path $N_1N_2$, followed by $N_1W_2$ and $W_1N_2$ with equal probability, while $W_1W_2$ is the least preferred option [Fig.~\ref{fig:fig3} b (ii)]. Similarly, during the return journey from the new nest to the old nest, leaders most frequently choose the path $N_2N_1$, followed by $W_2N_1$ and $N_2W_1$ equally, with $W_2W_1$ being the least preferred option [Fig.~\ref{fig:fig3} b (ii)]. The emergence of these preferences following the introduction of the bridge arises because the expected travel time between points $D_1$ and $D_2$ (and similarly between $D_2$ and $D_1$) is shorter for the paths $N_1N_2$ and $N_2N_1$ than for the alternative routes. 

\subsection{Exploration-exploitation theory}
\label{ex-ex-Theory}
\subsubsection{Exploration: Probability of reaching destination}
\label{ex-ex:exploration}
Suppose an ant starts from the origin to explore a path toward the destination.  Assume that at time step $t$, the ant has covered a distance $d$ units  from the origin toward the destination along a single path $\pi$. We want to compute the probability of subsequently reaching the destination, located at distance $l$ from the origin, without returning to the origin. Let this probability be denoted by $P_\pi(d)$. By definition, the boundary conditions are $P_\pi(0)=0$ and $P_\pi(l)=1$.
Following the microscopic unbiased random-walk dynamics, the hitting probability satisfies
\begin{equation}
	P_\pi(d) = \frac12 P_\pi(d+1) + \frac12 P_\pi(d-1).
\end{equation}
This linear difference equation has a linear solution of the form $P_\pi(d)=Ad+B$. Applying the boundary conditions at $d=0$ and $d=l$, we obtain 
\begin{equation}
P_\pi(d)=\frac{d}{l}.
\end{equation}
If an agent returns to the origin, it restarts the same process and continues until it reaches the destination. Therefore, the total probability of eventually reaching the destination is one. However, when multiple paths connect the origin to the destination, the probability of reaching the destination via a particular path may differ if the path lengths are unequal.

In the path network $P_0$ [see Fig.~\ref{fig:abm}(i)], the origin $D_1$ and destination $D_2$ are connected by two paths ($N_1W_2$ and $W_1N_2$) of equal lengths. Hence, it is straightforward to see that the probability of reaching the destination via either path is $\frac{1}{2}$.

The situation becomes more interesting for the path network $P_L$ [see Fig.~\ref{fig:abm}(ii)], where $D_2$ can be reached from $D_1$ (and vice versa) by four paths: $N_1W_2,\, W_1N_2,\, N_1N_2,$ and $W_1W_2$.
Since these paths have different lengths, calculating the probability of reaching $D_2$ from $D_1$ via a given path is non-trivial.

We first compute the probability of reaching the intermediate node $D_L$ from $D_1$, and then the probability of reaching $D_2$ from $D_L$. Assuming that the decisions at $D_1$ and $D_L$ are independent, the total probability can be obtained by multiplying these two contributions.

\textbf{From $D_1$ to $D_L$:}
 Let $q_{N_1}$ and $q_{W_1}$ denote the probabilities of eventually reaching $D_L$ via $N_1$ and $W_1$, respectively. It may be argued that
 \begin{equation}
 	q_{N_1} = \frac{1}{2} P_{N_1}(1) 
 	+ \left[1 - \frac{1}{2}P_{N_1}(1) - \frac{1}{2}P_{W_1}(1)\right] q_{N_1},\label{eq:qn1}
 \end{equation}
where the first term in the R.H.S. is the probability of reaching $D_L$ from $D_1$ via $N_1$ in a single attempt, without returning to the origin; the second term accounts for the event that after taking one step on either of the paths, the agent returns to the origin [probability is $1 - \frac{1}{2}P_{N_1}(1) - \frac{1}{2}P_{W_1}(1)$] and then eventually reaches the destination using $N_1$ (probability is $q_{N_1}$). Therefore, $q_{W_1}=1-q_{N_1}$. Since, $\frac{1}{2}P_{N_1}(d=1)=\frac{1}{2l_N}$ and 
$\frac{1}{2}P_{W_1}(d=1)=\frac{1}{2l_W}$, one finds that
\begin{eqnarray}
q_{N_1}=\frac{l_W}{l_N + l_W}, 
\quad
q_{W_1} = 1 - q_{N_1} = \frac{l_N}{l_N + l_W}.\label{eq:q1}
\end{eqnarray}

\textbf{From $D_L$ to $D_2$:}
Now consider the transport from $D_L$ to $D_2$ via $N_2$ and $W_2$. From $D_L$, an agent can choose among four segments---$N_1$, $W_1$, $N_2$, and $W_2$---with equal probability, $1/4$. However, choosing $N_1$ or $W_1$ does not lead to the destination; instead, the agent eventually returns to $D_L$ (since $q_{N_1}+q_{W_1}=1$). Along the same line of argument behind Eq.~(\ref{eq:qn1}), one has 
\begin{equation}
	q_{N_2} = \frac{1}{4} P_{N_2}(1) 
	+ \left[1 - \frac{1}{4}P_{N_2}(1) - \frac{1}{4}P_{W_2}(1)\right] q_{N_2},
\end{equation}
leading to 
\begin{eqnarray}
q_{N_2}=\frac{l_W}{l_N + l_W}, \quad q_{W_2} = \frac{l_N}{l_N + l_W}.\label{eq:q2}
\end{eqnarray}

\textbf{Composite paths:}
Assuming independence of decisions at $D_1$ and $D_L$, the probability of reaching $D_2$ from $D_1$ via a composite path $\alpha\beta$, where $\alpha \in \{N_1, W_1\}$ and $\beta \in \{N_2, W_2\}$, is
\begin{eqnarray}
q_{\alpha\beta} = q_\alpha \, q_\beta.\label{eq:qab}
\end{eqnarray}

Similarly, the probability of returning from $D_2$ to $D_1$ via the reverse path $\beta\alpha$ is given by
\begin{eqnarray}
q_{\beta\alpha} = q_\beta \, q_\alpha.
\end{eqnarray}

\subsubsection{Exploitation: Probability of reaching destination}
\label{ex-ex:exploitation}
We now incorporate exploitation. After discovering a path, suppose an ant reuses that path with probability $p$ (exploitation); otherwise, with probability $(1-p)$, it explores again. There is a set of four possible  paths $\mathcal{S}=\{N_1W_2, W_1N_2, N_1N_2, W_1W_2\}$. If $f_i$ denotes the probability of using path $i \in \mathcal{S}$ to reach the point $D_2$ from $D_1$, it must satisfy the self-consistency equation: $f_i = p f_i + (1-p) q_i$, which yields 
\begin{eqnarray}
f_i = q_i.\label{eq:f=q}
\end{eqnarray}

\subsubsection{Predicting choice-frequencies}
The choice-frequencies as estimated from experimental data and maximum-likelihood method (see 	Fig.~\ref{fig:fig3}) requires the knowledge of mean vector, $\bm{\mu}^a$ and covariance matrix, ${\sf \Sigma}^a_j$. 

Before we provide the relevant calculations below, a remark is in order: The choice frequency vector ${\bm x}^{a}_{ij} = \left(x^{a,k}_{N_1W_2},\, x^{a,k}_{W_1N_2},\, x^{a,k}_{N_1N_2},\, x^{a,k}_{W_1W_2}\right)$, where $a \in \{tl, rl\}$, has four components; however, only three of them are independent since the sum of four components is unity. Therefore, the multivatiate Gaussian distribution of ${\bm x}^{a}_{ij}$ is completely characterized by only three components of ${\bm \mu^a}$ and nine elements  (say, $[{\sf \Sigma}^a_j]_{ss'}$, where $s,s'\in\{N_1W_2, W_1N_2, N_1N_2\}$) of the covariance matrix. ${\sf \Sigma}^a_j$.

\textbf{Mean vector $\bm{\mu}^a$:}
We first compute the mean vector $\bm{\mu}^a$, where $a \in \{tl, rl\}$. By definition,
\begin{equation}
	\bm{\mu}^a = \langle \bm{x}^a_{ij} \rangle 
	= \frac{1}{\tau^a_j} \sum_{k=1}^{\tau^a_j} \langle \bm{\theta}^{a,k}_{ij} \rangle,
\end{equation}
where $\tau^{rl}_j = T_j - 1$, $\tau^{tl}_j = T_j$, and the angular bracket denotes the average over all the realizations of the corresponding random variable. Here, $\bm{\theta}^{a,k}_{ij}$ is a random vector generated from the process described in Appendix~\ref{ex-ex:exploration} and \ref{ex-ex:exploitation}. Therefore,
\begin{eqnarray}
\langle \bm{\theta}^{a,k}_{ij} \rangle&=& (f_{N_1W_2},\, f_{W_1N_2},\, f_{N_1N_2},\, f_{W_1W_2}), \nonumber\\
&=& (q_{N_1W_2},\, q_{W_1N_2},\, q_{N_1N_2},\, q_{W_1W_2}),
\end{eqnarray}
where we have used Eq.~(\ref{eq:f=q}). Finally, invoking Eqs.~(\ref{eq:q1}), (\ref{eq:q2}), and (\ref{eq:qab}), determines $\bm{\mu}^a$. In the experiments, we use $l_N \approx 20$ cm and $l_W \approx 48$ cm.

\textbf{Covariance matrix ${\sf \Sigma}^a_j$:}
Next, we compute the covariance matrix ${\sf \Sigma}^a_j$. By definition, the $(s,s')$-th element is
\begin{align}
	[{\sf \Sigma}^a_j]_{ss'} 
	&= \mathrm{Cov}\bigl([\bm{x}^a_{ij}]_s,\,[\bm{x}^a_{ij}]_{s'}\bigr)\nonumber\\
	&= \frac{1}{(\tau^a_j)^2} \sum_{k=1}^{\tau^a_j} \sum_{k'=1}^{\tau^a_j} 
	\mathrm{Cov}\bigl(\theta^{a,k}_s,\, \theta^{a,k'}_{s'}\bigr).
\end{align}

This can be decomposed as
\begin{align}
	[\Sigma^a_j]_{ss'} 
	&= \frac{1}{(\tau^a_j)^2} \sum_{k=1}^{\tau^a_j} 
	\mathrm{Cov}\bigl(\theta^{a,k}_s,\, \theta^{a,k}_{s'}\bigr)\nonumber\\
	&+ \frac{2}{(\tau^a_j)^2} \sum_{n=1}^{\tau^a_j-1} (\tau^a_j - n)\,
	\mathrm{Cov}\bigl(\theta^{a,k}_s,\, \theta^{a,k+n}_{s'}\bigr).\label{eq:c6}
\end{align}

Let us now find the explicit expressions of the covariances:

(i) \textbf{Same-step covariance:}
\begin{equation}
	\mathrm{Cov}\bigl(\theta^{a,k}_s,\, \theta^{a,k}_{s'}\bigr)
	= \langle \theta^{a,k}_s \theta^{a,k}_{s'} \rangle 
	- \langle \theta^{a,k}_s \rangle \langle \theta^{a,k}_{s'} \rangle.
\end{equation}

Since exactly one path is chosen at each step and $\theta^{a,k}_s \in \{0,1\}$, we have
$\theta^{a,k}_s \theta^{a,k}_{s'} = 0$ for $s \neq s'$ and  $\theta^{a,k}_s \theta^{a,k}_{s} = \theta^{a,k}_{s}$. 
Therefore, $\theta^{a,k}_s \theta^{a,k}_{s'} = \delta_{ss'}\, \theta^{a,k}_s.$ Here and henceforth, $\delta$ is Kronecker delta.
Taking expectation, we obtain $\langle \theta^{a,k}_s \theta^{a,k}_{s'} \rangle = \delta_{ss'}\, f_s,$ and hence
\begin{equation}
	\mathrm{Cov}\bigl(\theta^{a,k}_s,\, \theta^{a,k}_{s'}\bigr)
	= \delta_{ss'} q_s - q_s q_{s'}.\label{eq:c8}
\end{equation}
Here and henceforth, we do not distinguish between $f$ and $q$ in the light of Eq.~(\ref{eq:f=q}).

(ii) \textbf{Cross-step covariance:}
\begin{equation}
	\mathrm{Cov}\bigl(\theta^{a,k}_s,\, \theta^{a,k+n}_{s'}\bigr)
	= \langle \theta^{a,k}_s \theta^{a,k+n}_{s'} \rangle 
	- q_s q_{s'}.\label{eq:c9}
\end{equation}

The first term in R.H.S. of Eq.~(\ref{eq:c9}) can be written as
\begin{eqnarray}
	\langle \theta^{a,k}_s \theta^{a,k+n}_{s'} \rangle
	&=& \sum_{\theta^{a,k}_s}[ \theta^{a,k}_s \, \mathrm{Prob}(\theta^{a,k}_s) \times \nonumber\\
	&&\sum_{\theta^{a,k+n}_{s'}} \theta^{a,k+n}_{s'} \,
	\mathrm{Prob}(\theta^{a,k+n}_{s'} \mid \theta^{a,k}_s)].\qquad\quad\label{eq:nsc}
\end{eqnarray}

 Motivated by a reasoning given in the following paragraph, we propose the following ansatz for further evaluation of Eq.~(\ref{eq:nsc}):
\begin{equation}
	\mathrm{Prob}(\theta^{a,k+n}_{s'} \mid \theta^{a,k}_s)
	= p^n \delta_{ss'} + (1 - p^n)\, q_{s'},\label{eq:pnn}
\end{equation}
which will be proven to be true subsequently using the method of induction.

The case of $n=1$ is rather easy to prove: With probability $p$ the agent reuses the same path $s$; otherwise, with probability $(1-p)$, it explores and selects a path according to the discovery distribution, choosing $s'$ with probability $q_{s'}$. Therefore,
\begin{equation}
	\mathrm{Prob}(\theta^{a,k+1}_{s'} \mid \theta^{a,k}_s)
	= p\, \delta_{ss'} + (1-p)\, q_{s'}.
	\label{one step-correlation}
\end{equation}
At each step, the agent either repeats its previous path with probability $p$, or explores with probability $(1-p)$ and selects a path according to exploration probabilities (see Appendix~\ref{ex-ex:exploration}). Therefore, over $n$ steps, the agent continues to use the same initial path $s$ only if it chooses to repeat at every step, which occurs with probability $p^n$, yielding $s' = s$. On the other hand, if the agent explores at least once within these $n$ steps (which occurs with probability $1 - p^n$), then its path choice at step $k+n$ is governed by the exploration rule. Hence, Eq.~(\ref{eq:pnn}) is intuited.

Now to see that Eq.~(\ref{eq:pnn}) is indeed true for all $n\in\mathbb{N}$, we assume that  Eq.~(\ref{eq:pnn}) holds for $n=m$. Then,
\begin{eqnarray}
	\mathrm{Prob}(\theta^{a,k+m+1}_{s'} \mid \theta^{a,k}_s)
	&= \sum_r \mathrm{Prob}(\theta^{a,k+m+1}_{s'} \mid \theta^{a,k+m}_r)\times \nonumber \\
	&~~~~~~~\mathrm{Prob}(\theta^{a,k+m}_r \mid \theta^{a,k}_s)\nonumber\\
	&= \sum_r \bigl[p \delta_{rs'} + (1-p) q_{s'}\bigr] \times \nonumber\\
	 &~~~~~~~\bigl[p^m \delta_{rs} + (1 - p^m) q_r\bigr]\nonumber\\
	&= p^{m+1} \delta_{ss'} + (1 - p^{m+1}) q_{s'},\nonumber\\\label{eq:p|}
\end{eqnarray}
implying that Eq.~(\ref{eq:pnn}) holds for $n=m+1$. We already know that  Eq.~(\ref{eq:pnn}) holds good for $n=1$ (see Eq.~\ref{one step-correlation}). In conclusion, Eq.~(\ref{eq:pnn}) must be true for all $n$.

Putting Eq.~(\ref{eq:p|}) in Eq.~(\ref{eq:nsc}) and subsequently, substituting  resulting form of Eq.~(\ref{eq:nsc}) in Eq.~\ref{eq:c9}, we arrive at
\begin{equation}
	\mathrm{Cov}\bigl(\theta^{a,k}_s,\, \theta^{a,k+n}_{s'}\bigr)
	= p^n \bigl(\delta_{ss'} q_s - q_s q_{s'}\bigr).\label{eq:c14}
\end{equation}

Finally, putting Eq.~(\ref{eq:c9}) and Eq.~(\ref{eq:c14}) in Eq.~(\ref{eq:c6}), one gets the compact expression of the covariance matrix in question:
\begin{equation}
	[{\sf \Sigma}^a_j]_{ss'} 
	= (\delta_{ss'} q_s - q_s q_{s'}) 
	\frac{1}{(\tau^a_j)^2}
	\left[
	\tau^a_j + 2 \sum_{n=1}^{\tau^a_j - 1} (\tau^a_j - n)\, p^n
	\right].
\end{equation}

\end{document}